\documentclass[aip,jap]{revtex4-1}

\usepackage[sort&compress]{natbib}
\usepackage{amsmath} 
\usepackage{graphicx}

\graphicspath{{./GraphsJAP/}} 

\begin{document}

\title{Super-Resolving Quantum Radar: Coherent-State Sources with Homodyne Detection Suffice to Beat the Diffraction Limit 
}

\author{Kebei Jiang}
\affiliation{Hearne Institute for Theoretical Physics and Department of Physics and Astronomy \\
Louisiana State University, Baton Rouge, LA 70803, USA}
\author{Hwang Lee}
\affiliation{Hearne Institute for Theoretical Physics and Department of Physics and Astronomy \\
Louisiana State University, Baton Rouge, LA 70803, USA}
\author{Christopher C.\ Gerry}
\affiliation{Department of Physics and Astronomy, Lehman College, The City University of New York, Bronx, New York 10468-1589, USA }
\author{Jonathan P.\ Dowling}
\email{jdowling@lsu.edu}
\affiliation{Hearne Institute for Theoretical Physics and Department of Physics and Astronomy \\
Louisiana State University, Baton Rouge, LA 70803, USA}
\affiliation{Beijing Computational Science Research Center, Beijing 100084, China }

\date{\today}

\begin{abstract}

There has been much recent interest in quantum metrology for applications to sub-Raleigh ranging and remote sensing such as in quantum radar. For quantum radar, atmospheric absorption and diffraction rapidly degrades any actively transmitted quantum states of light, such as N00N states, so that for this high-loss regime the optimal strategy is to transmit coherent states of light, which suffer no worse loss than the linear Beer's law for classical radar attenuation, and which provide sensitivity at the shot-noise limit in the returned power. We show that coherent radar radiation sources, coupled with a quantum homodyne detection scheme, provide both longitudinal and angular super-resolution much below the Rayleigh diffraction limit, with sensitivity at shot-noise in terms of the detected photon power. Our approach provides a template for the development of a complete super-resolving quantum radar system with currently available technology. 

\end{abstract}

\maketitle


\section{Introduction}
\label{sec:intro}

\begin{figure}
	\centering
		 \includegraphics[width=1\textwidth]{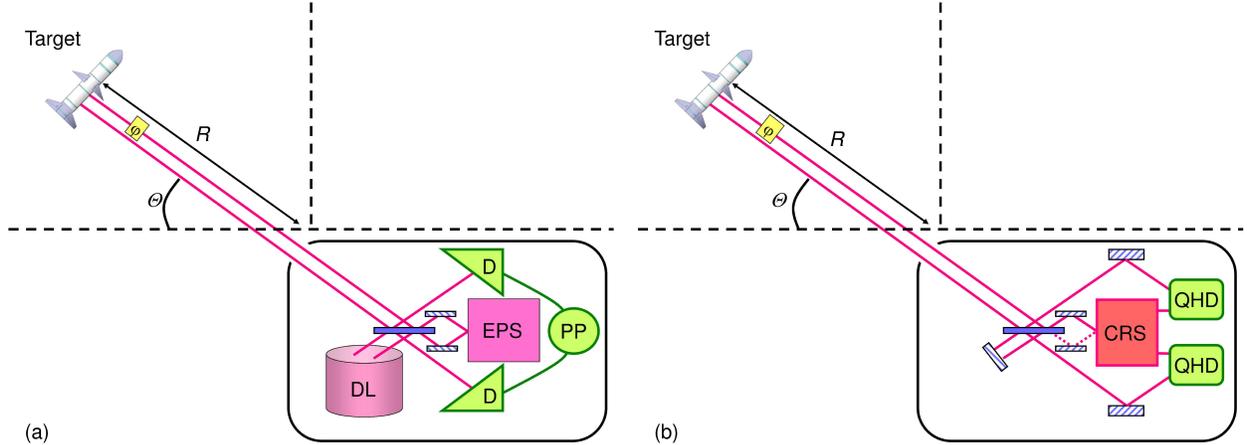}
	\caption{Here we compare two monostatic quantum radar systems in a Michelson configuration. The first (a) uses an entangled photon source and photon number resolving detectors, and the second (b) uses a coherent radar source and quantum homodyne detection. In both the range distance $R$ and the altitudinal angle above the horizon $\varTheta$ are shown. (For clarity the azimuthal angle $\varPhi$ along the horizon is suppressed.) The phase shifter $\varphi$ is an icon for the signal, the relative phase between the two arms, which carries the range information. In (a) half of the entangled state from the entangled photon source (EPS) is reflected off the target and half is stored in a delay line (DL), and the photon-counting detectors (D) that send their data to a post-processor (PP). In (b) the entangled source is replaced with a coherent radar source (CRS), the delay line is replaced with a mirror, and the detection is carried out using two quantum homodyne detectors (QHD). Dotted red lines in (b) indicate the quantum vacuum that enters in the unused port of the interferometer. The solid blue rectangle is a beam splitter and the striped blue rectangles are mirrors. The red lines connecting the CRS to the QHDs shows that the same radar source may be used as the local oscillator in the homodyne detection.}
	\label{FIG01}
\end{figure}

Ever since work of the Louisiana State University (LSU) group on quantum lithography, they have been able to show that entangled quantum states of the electromagnetic field, such as the Schr$\ddot{\text{o}}$dinger-cat-like N00N states of the form $\vert N::0\rangle_{AB}^{\varphi}=(\vert 0\rangle_{\text{A}} \vert N\rangle_{\text{B}}+ e^{\mathrm{i} N \varphi} \vert N\rangle_{\text{A}} \vert 0\rangle_{\text{B}})/\sqrt{2}$, provide a resolving power that is sub-Rayleigh diffraction limited (super-resolution) and a sensitivity that is sub-shot-noise limited (super-sensitivity)\cite{Boto2000,Kok2001, Kok2004, Dowling2008}. Here $N$ is the photon number and A and B label the two modes or ‘arms’ of either a Mach-Zehnder or Michelson interferometer, the latter of which is typically deployed in coherent lidar and radar systems. (See Fig.~\ref{FIG01} and \ref{FIG02}.) This realization of a quantum-entanglement advantage led to proposals to develop remote quantum sensors, quantum lidar and quantum radar in particular, where such quantum states of light are actively transmitted through the atmosphere, reflected off the target, and then upon return provide sub-Rayleigh diffraction resolution in ranging \cite{Lanzagorta2012}.

Notwithstanding claims to the contrary \cite{Allen2005}, to exploit N00N states in remote sensing they must be deployed in an interferometric mode, such as used in classical coherent lidar and radar systems, wherein half of the N00N state is transmitted to the target and back, whereupon that half is mixed with the other half at the detector. We illustrate such a quantum-entangled radar system in Fig.~\ref{FIG01}(a) where the entangled photon source is attached to a Michelson interferometer in a monostatic configuration. (The term `monostatic' means the source and the detector are at the same location as compared to `bistatic' where they are in different locations.) Such a quantum-entangled radar scheme requires that the half of the N00N state be retained at the radar station and stored in a low-loss delay line for a time equal to the round trip time that the other half takes between the radar station and the target. Implementing the correct delay time then would require at least approximate advanced knowledge of the distance to the target, and such an entangled-photon quantum radar system could not be a stand alone system, but rather would have to provide improved ranging in conjunction with the simultaneous deployment of a conventional radar at the same site.  

An additional immediate problem arose with this entangled-photon approach to quantum radar when two different groups pointed out in 2007 and 2008 that the N00N states are very susceptible to the linear loss expected from absorption and scattering in the atmosphere and also from diffraction (the last of which would imply that not all of the N00N state would be detected upon return due to the finite size of the detector aperture) \cite{Rubin2007, Gilbert2008a, Gilbert2008b}. Unavoidable loss would also be present in the delay line. The LSU group was recently able to provide a quantum theoretical interpretation of this ‘super-Beer’ loss behavior for N00N states in terms of the quantum ‘which-path’ information available to the environment upon photon loss \cite{Huver2008, Knysh2011}. In that same paper they provided an intuitive solution to partially mitigate loss by replacing N00N states with the so-called M$\&$M$'$ states, which have the form $\vert M::M'\rangle_{AB}^{\varphi}=(\vert M\rangle_{\text{A}} \vert M'\rangle_{\text{B}}+ e^{\mathrm{i} (M-M') \varphi} \vert M'\rangle_{\text{A}} \vert M\rangle_{\text{B}})/\sqrt{2}$. In such states, if $M-M'=N$, then the state remains $N$-fold super-resolving, and for low loss can still do better than shot-noise in sensitivity. In such M$\&$M$'$ states the vacuum-number state in the N00N state is replaced with a low-number Fock state that acts as a ‘decoy’ photon. This decoy photon acts to keep complete which-path information from becoming available to the environment with just a single photon lost, and thence their approach — at least for a while — staves off decoherence, destruction of the entanglement, loss of phase information, and thence preserves the ranging information. 

At that junction the LSU group set out to numerically search Hilbert space for quantum states of the electromagnetic field that had the best sensitivity for a specified loss \cite{Lee2009}. Their results indicate that for the low loss regime the N00N states are optimal, for intermediate loss the M$\&$M$'$ states are optimal, but that \textit{at high loss only coherent states are optimal}. Konrad Banaszek, Ian Walmsley, and their collaborators, independently discovered this result about the same time \cite{Dorner2009, Demkowicz-Dobrzanski2009}. Since coherent states are the natural output of a conventional lidar or radar source, their conclusion was that switching from coherent to entangled states would not give sub-shot-noise sensitivity in the high-loss regime, since it is known that coherent states by themselves will always achieve at best shot-noise sensitivity in the return power \cite{Caves1981}. That is, \textit{super-sensitivity with entangled state transmission is impossible to achieve when total loss exceeds about 6 dB}. That means that that entangled-photon radar is useless for most applications. This conclusion left open the question of using coherent states for \textit{super-resolving} lidar and radar in the high-loss regime, while still operating at the shot-noise limit in sensitivity with respect to the return power. That is can we remove the fragile entanglement and still beat the Rayleigh diffraction limit with coherent states alone? The answer is, yes we can.  

In 2007 the group of Andrew White demonstrated that coherent states could indeed provide super-resolution if a quantum detection scheme was deployed \cite{Resch2007}. In this scheme of Resch \emph{et al.} they projected the return coherent state onto a N00N-state basis to extract resolution for a particular N00N state component of the two-mode coherent field in the interferometer by doing high-efficiency $N$-photon counting \cite{Gao2010}. Using this technique they demonstrated six-fold super resolution. However such a scheme throws away almost all of the returning photons and is also much worse than shot-noise in sensitivity. Hence the scheme of Resch \emph{et al.} is far from ideal in situations where only few photons are expected to return from a distant target as is typical for long-range radar systems. In addition such a scheme would require photon number resolving detectors of high efficiency that are not currently available at lidar and radar wavelengths.

The work of Resch \emph{et al.} motivated the LSU group to seek a coherent-state quantum detection scheme that was both super-resolving and that operated at the shot-noise limit in sensitivity, which is provably the best sensitivity you can achieve with coherent states alone \cite{Caves1981}, and which would also preserve the information in all the returning photons. They found such a scheme in 2010 and showed that super-resolution (in longitudinal ranging) at the shot-noise limit with coherent states can be achieved using quantum parity detection, a measurement that detects if the number of photons exiting one port of the interferometer is either or even or odd \cite{Gao2010}. Christopher Gerry and his collaborators have particularly championed the photon-number parity measurement in quantum metrology \cite{Bollinger1996, Gerry2000, Campos2003, Hofmann2009, *Gerry2010, *Chiruvelli2011, *Seshadreesan2013}. Hwang Lee and Yang Gao in the LSU group have shown that parity detection provides a unifying measurement scheme in quantum metrology, in that it provides sub-shot-noise sensitivity with respect to a wide range of quantum-entangled states of the electromagnetic field \cite{Gao2008}. In all cases the LSU group has investigated so far, parity detection always saturates the quantum Cramer-Rao bound, which is the lower bound on the best sensitivity in any quantum metrology scheme \cite{Anisimov2010}. 

One remaining practical drawback to that 2010 proposal, for super-resolving lidar and radar ranging with coherent states alone, was that back then the way to implement the quantum parity detection involved either the use of very strong Kerr nonlinearities \cite{Gerry2002} or high-efficiency photon-number resolving detectors \cite{Gao2010}. The Kerr approach would require high-power in the return radar field, which is not usually the case for many applications of interest. For long-wavelength radar systems, due to diffraction, the source emits power in nearly a spherical wave, as does the reflecting target, and so the ratio of power transmitted to power received back scales as $1/R^{4}$, where $R$ is the range distance to the target, which is typically tens to hundreds of kilometers. For example, if the target is 100 kilometers distant, then for a radar system that transmits a kilowatt in outgoing power, the return power will only be about 400 picowatts (assuming that both the emitter and target have a cross sectional area of a square meter). Alternatively, the photon-number resolved detection approach is problematic in the infrared and particularly the microwave and longer wavelength radar regimes where such photon-number detectors have extremely low efficiency and where thus the parity detection advantage would be lost \cite{Rosenberg2005}. 

The breakthrough that allows us now to apply this parity measurement technique to lidar and radar wavelengths came in 2010 when, in collaboration with Girish Agarwal, the LSU group were able to show that parity detection of the coherent state can be carried out with a simple quantum homodyne detection scheme \cite{Plick2010}. Homodyne detection is a process, by which the return radar state in the system is mixed with a local oscillator, which itself is a stable radar source of the same frequency as the transmitted beam and with a known and adjustable phase \cite{Lvovsky2009}. While quantum homodyne is a standard technique in quantum optics it has its origins in a World War II radar technology known as the balanced-mixer radiometer \cite{Dicke1946}. In this language the ‘balanced mixer’ is what we in quantum optics call the 50-50 beam splitter, and we will use the latter terminology throughout. The point was to recognize the well-known fact that a parity measurement of a single-mode electromagnetic field is equivalent to the measurement of the field's Wigner function, a quantum optical phase-space distribution of the electromagnetic field, at the origin in phase space. In the quantum optics community it is well established that the complete Wigner function may be reconstructed experimentally through a process called quantum tomography, which is implemented through the collection of many different phase-sensitive homodyne measurements of the field \cite{Lvovsky2009}. However the process of quantum homodyne tomography can be time consuming and resource intensive if the full Wigner function is to be obtained at high fidelity, but nevertheless it yields parity, as noted by Campos and Gerry \cite{Campos2003}. 

However we do not need the complete Wigner function here but only its value at the origin in phase space \cite{Cahill1969, Royer1977, Banaszek1999}. That realization greatly simplifies the number of homodyne measurements needed from hundreds to only one or two \cite{Campos2003, Plick2010}. Hence parity detection can be carried out using commercial-off-the-shelf homodyne radar detection components that are arranged to implement this quantum detection scheme. Since the radar source is also a commercial-off-the-shelf radar emitter, \textit{our proposed super-resolving quantum radar system may be constructed with current technology with minimal reconfiguration of the existing radar system}. An additional advantage of coherent-state quantum radar is that the need for the delay line, required in the entangled state protocol, disappears. The delay line may be replaced with a stable radar local oscillator and thus one does not need advance knowledge of the distance to the target. Thus our proposed quantum radar system then operates at as a stand-alone system. The homodyne technique has the additional advantage that the local oscillator that is mixed with the signal acts as an amplifier to boost the signal well above the thermal noise floor of radar detectors. One still needs high-efficiency detectors (which can be routinely made with superconductor technology) but not the number resolving feature. With such a scheme the presence or absence of a even a single photon in the return field may be detected \cite{Leonhardt1997}.

In their 2010 paper, in collaboration with Jerome Luine at Northrop-Grumman, the LSU compared parity measurement with a closely related ‘on-off’ photon detection scheme, where the latter is a scheme in which the detector distinguishes between zero photons versus more than zero in a single mode. In their numerical simulation they showed that the on-off scheme provides super-resolution comparable to that of parity detection \cite{Gao2010}. Recently the group of Ulrik Andersen has proposed and implemented a homodyne version of the on-off detection scheme and has experimentally demonstrated, for the first time ever, super-resolution at the shot-noise limit with coherent states and a quantum detection scheme \cite{Distante2013}. Independently the group of Hagai Eisenberg has also experimentally demonstrated super-resolution at the shot-noise limit using only coherent light and homodyne-based parity detection \cite{Eisenberg2012}. These results, both carried out with visible light, vindicate the LSU group's approach and lend credence to the notion that the entire scheme may now be scaled to the infrared, microwave, and long-wavelength radar regimes. 

One final thing only now remains to be done to apply coherent states and parity measurement to a practical quantum radar system. In all the schemes discussed above we have shown how to obtain super-resolved ranging information, which is distance to the target $R$. To completely characterize the location of a distant target we need in addition super-resolving altitudinal angle $\varTheta$ (location of the target above the horizon) and azimuthal angle $\varPhi$ (location of the target along the horizon). (See Fig.~\ref{FIG01}.) This angular information is of particular importance in long-wavelength radar systems, where as noted above, the target may be tens or hundreds of kilometers distance and the radar power is reflected back in a spherical wave. That means that the return radar signal received at the detector arrives essentially as a plane wave, which makes the determination of the angular position of the target most difficult. The trick is to convert the angular information into a phase shift and then use the same technique used to acquire super-resolved range information to also obtain super-resolved angular information. Thus in this paper we show, for the first time, how a quantum homodyne-based parity measurement scheme can provide angular super resolution, in addition to ranging super resolution, still while deploying only coherent states of the radar field. Together with ranging we then have complete super-resolved angular location of the target embodied in the determination of $R$, $\varTheta$ and $\varPhi$. Since many targets of interest are moving at high velocity --- Mach 30 is not uncommon for some applications --- by carrying out time differentiation on these three parameters, super-resolved velocity and acceleration information of the target may be also obtained. This procedure gives us the ability to predict the target’s future trajectory with great precision, particularly if the target is moving ballistically. 

In Section II we will review the use of quantum parity detection to produce super-resolving ranging measurements at the shot-noise limit. In Section III will discuss parity detection’s implementation as a quantum homodyne detection scheme. In Section IV we will present our newest result that shows how to modify the technology to provide super-resolved angular determination. In Section V we will conclude and summarize.

\section{SUPER-RESOLVED RANGING WITH PARITY DETECTION}
\begin{figure}
	\centering
		 \includegraphics[width=0.5\textwidth]{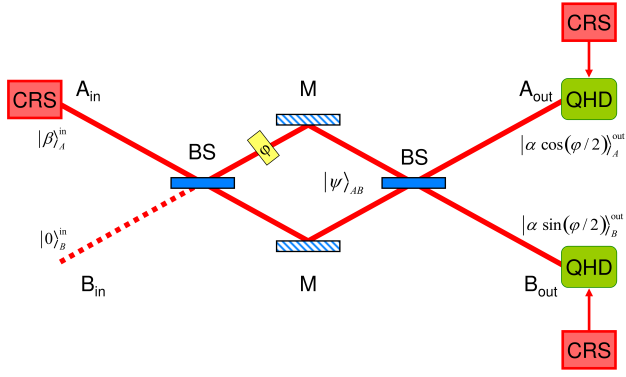}
	\caption{Here for simplicity we show the monostatic Michelson interferometer in Fig.~\ref{FIG01}(b) unfolded into an equivalent bistatic Mach-Zehnder configuration. The coherent state from the coherent radar source (CRS) is incident in upper mode A and vacuum in lower mode B at the left. After the first beam splitter transformation (BS) we have a two-mode coherent state. After the phase shifter $\varphi$, which encodes the range $R$, this state becomes the two-mode coherent state of with a relative phase difference that is reflected off the mirrors (M). Finally after the final beam splitter (BS) at the right, we have the attenuated coherent state with the phase information and we implement the parity operator measurement in both the upper mode A and the lower mode B via quantum homodyne detection (QHD). The same CRS is also used to feed both QHDs to implement the balanced homodyne procedure. (In the monostatic folded Michelson interferometer the CRS is located at the input and at the output facilitating this operation.)}
	\label{FIG02}
\end{figure}

It is typical to analyze the Michelson interferometer (MI) in Fig.~\ref{FIG01}(b) in an unfolded Mach-Zehnder interferometer (MZI) shown in Fig.~\ref{FIG02}. The performance of an MZI and the MI are identical. The MZI corresponds to a bistatic radar system where the source and detector are at different locations and the MI to a monostatic system where the source and detector are co-located. The only physical difference is that the MZI has two separate beam splitters (BS) and the MI has a single beam splitter that is utilized twice; once upon emission and again upon detection. As shown in Fig.~\ref{FIG02} the unknown phase to be detected is denoted by $\varphi$ and is the given by $\varphi = k \ell$, where $k$ is the wave number (with $\lambda$ the wavelength) and $\ell$ is the path-length \textit{difference} between the two arms of the interferometer that lie between the two BS. In the MI configuration of Fig.~\ref{FIG01}(b) the length of the lower reference arm can be made to be zero and thence $\ell = R$ the sought-after range to the target.

In quantum optics the output of a coherent electromagnetic source is called a coherent state $\vert \alpha \rangle$ where $\alpha = \sqrt{\bar{n}} e^{\mathrm{i} \theta}$ is a complex number or phasor encoding information about both the amplitude $\sqrt{\bar{n}}$ and the phase $\theta$ of the field. In this notation $\vert \alpha \vert^2 = \bar{n}$  is the average number of photons in the field that is proportional to the return intensity. Since we are only interested in the phase difference accumulated upon propagation, without loss of generality we may take $\theta = 0$ and so $\alpha = \sqrt{\bar{n}}$ is real. Since $\vert \alpha \rangle$ is a quantum mechanical state the intensity can never be measured precisely and for such a state a measurement of the intensity will yield only $\bar{n}$  photons on average with quantum fluctuations about this mean on the order of $\Delta n = \sqrt{\bar{n}}$. Only in the classical limit, when the intensity is very large, is $\Delta n$  negligible compared to $\sqrt{\bar{n}}$ and we can then treat the system using classical radar theory in that approximation. As we are particularly interested in the situation where very few photons return to the detector, in that regime  $\Delta n$ cannot be neglected compared to $\bar{n}$  and the quantum theory of light must be used \cite{Gerry2005a}. 

\begin{figure}[h]
	\centering
		 \includegraphics[width=0.4\textwidth]{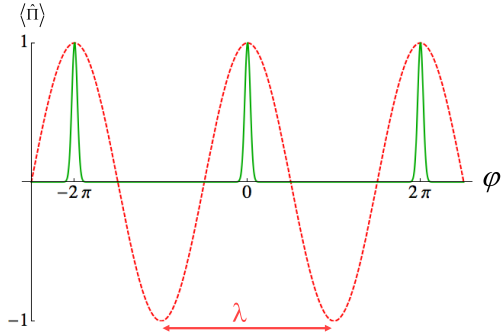}
	\caption{Here we show the signal of quantum parity detection (green solid curve) against the ordinary classical signal obtained by differencing the intensity of the two detectors and scaled by the intensity (red dashed curve). The parity curve is for a return power of $\bar{n}=100$ giving a ten-fold improvement in the fringe resolution compared to that of the classical curve where the peak-to-peak spacing is at the diffraction limit of $\lambda$.}
	\label{FIG03}
\end{figure}

As shown in Fig.~\ref{FIG02} we emit a coherent state input mixed with vacuum at the first BS that we write as $\vert \beta \rangle_{\text{A}}^{\text{in}} \vert 0 \rangle_{\text{B}}^{\text{in}}$. After traversing the interferometer the output state becomes $\vert \alpha \cos (\varphi/2) \rangle_{\text{A}}^{\text{out}} \vert \alpha \sin (\varphi/2) \rangle_{\text{B}}^{\text{out}}$ \cite{Dowling2008}. Here $\alpha = e^{-\Gamma R/2} \beta$ is the attenuated coherent state at the detector, where $\Gamma$ is the linear intensity attenuation coefficient and $R$ is the range, and we assume the lower path is attenuation free to be consistent with the monostatic configuration of the Michelson operation where the reference path is small. We wish to implement parity measurement at output port B. We can compute the result by noting that the expectation of the parity operator is proportional to the Wigner function of the output state evaluated at the origin in phase space, 
\begin {align}
\langle \hat{\mathrm{\Pi}} \rangle = \langle e^{\mathrm{i}\pi \hat{n}} \rangle = \frac{\pi}{2} W(0,0),
\label{EQN01}
\end {align}
where $\hat{n}=\hat{b}^\dagger \hat{b}$ is the number operator and  $\hat{b}^\dagger$ is the mode operator for $\text{B}_{\text{out}}$ \cite{Plick2010}. The Wigner function for a coherent state has a particularly simple form \cite{Gerry2005b}, 
\begin {align}
W_{\alpha_{\varphi}}(\gamma,\gamma^*) = \frac{2}{\pi} \exp(-2\vert \gamma-\alpha_{\varphi} \vert^2),
\label{EQN02}
\end {align}
where $\gamma$ is a complex phase space coordinate and $\alpha_{\varphi}=\alpha \sin (\varphi /2)$  is the output coherent state in the mode $\text{B}_{\text{out}}$. The corresponding radar intensity in the mode is proportional to the mean photon number, defined by $\bar{n}_{\varphi}=\vert \alpha_{\varphi} \vert^2 = \bar{n} \sin^2(\varphi/2)$ . Combining these two equations we get,
\begin {align}
\langle \alpha_{\varphi} \vert \hat{\mathrm{\Pi}} \vert \alpha_\varphi \rangle = \exp \left(-2\vert \alpha_{\varphi} \vert^2 \right)
=\exp \left( -2 \bar{n} \sin^2 (\varphi/2) \right),
\label{EQN03}
\end {align}
which is the previous result of the LSU group\cite{Gao2010}. The on-off detection scheme has a similar form \cite{Distante2013}. This signal is plotted in Fig.~\ref{FIG03} as a solid curve with the dashed curve the signal from ordinary output intensity differencing (scaled by $\bar{n}$). It is clear that the parity signal is super-resolving. As shown in Ref.~\cite{Gao2010} the width of the parity central peak can be estimated by taking $\varphi \cong 0$ so that Eq. (3) becomes the Gaussian $\exp(-\bar{n}\varphi^2/2)$ so $\delta
\varphi =1/\sqrt{\bar{n}}$, where $\bar{n}$ is proportional to the return power. Converting this to an uncertainty in range we get,
\begin{align}
\delta R_{\text{Q}} = \frac{\lambda}{2 \pi \sqrt{\bar{n}}}, 
\label{EQN04}
\end{align}  
where the classical Rayleigh resolution would instead be just $\delta R_{\text{C}} = \lambda$ and so we are a factor of $2 \pi \sqrt{\bar{n}}$ below Rayleigh. For $\bar{n}=100$ average return photons the quantum result is 60 times smaller than the classical diffraction limit. \textit{Notice that by uncertainty we mean the full width at half maximum of the parity signal in Fig.~\ref{FIG03}, which is not the same as the sensitivity of phase estimation. The similar scaling of our superresolution and the general shot-noise limit is merely a coincidence.}

Since extraction of the parity signal will require some post-processing, one approach would be to simply measure the output intensity $\bar{n}_{\varphi}$ directly and plug the result into Eq.~\eqref{EQN03}. This approach is problematic for radar, in particular at low return photon numbers, since the best result would be obtained with efficient, low-noise, photon-number counters that are difficult to obtain at such long wavelengths. In addition the signal for few return photons will be well below the thermal noise floor of most detector at these frequencies. Also, in addition the quantum intensity fluctuations, $\Delta n = \sqrt{\bar{n}}$, there will be classical fluctuations due to instabilities in the radar source, turbulence in the atmosphere, and so forth. This is why intensity differencing is usually done between output modes $\text{A}_{\text{out}}$ and $\text{B}_{\text{out}}$ to give a common mode noise cancellation of these classical fluctuations. To obtain the optimal performance we should measure parity at each output using balanced quantum homodyne detection to extract the Wigner function at the origin directly and amplify the signal to well above the thermal noise floor. We discuss this approach in the next section.

\section{PARITY IMPLEMENTED WITH QUANTUM HOMODYNE DETECTION}

\begin{figure}
	\centering
		 \includegraphics[width=0.4\textwidth]{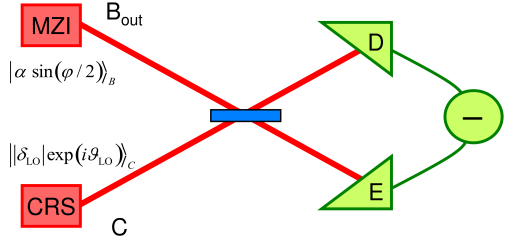}
	\caption{Here we depict the quantum homodyne detection. The lower mode output B of the Michelson interferometer (MI, unfolded and monostatic) or Mach-Zehnder interferometer (MZI, folded and bistatic) from Fig.~\ref{FIG02} is fed into a 50-50 beamsplitter. At the other input port C to this beam splitter we insert a strong coherent state $\big\vert \delta_\text{LO} \big\rangle_{\text{C}} = \big\vert \vert \delta_\text{LO} \vert \exp{(\mathrm{i} \vartheta_\text{LO})} \big\rangle_{\text{C}}$ with a known phase $\vartheta_{\text{LO}}$. After the beam splitter we carry out intensity differencing between the two detectors (D, E). The data is then inverted to extract the parity measurement. }
	\label{FIG04}
\end{figure}

We propose to carry out balanced homodyne detection of the parity signal at each of the two output ports of the interferometer. Such a detection scheme is shown in Fig.~\ref{FIG04}. As noted in the introduction, quantum homodyne detection was actually inspired by classical microwave radar technology called the balanced-mixer radiometer \cite{Dicke1946}. The balanced mixer is what we call here a beamsplitter. For this technique to work well in the microwave regime, at the quantum level, we also require that the detectors have a high quantum efficiency, which is routinely obtained these days with superconducting technology \cite{Stevenson2008}. We first work with the lower port $\text{B}_{\text{out}}$. The signal output for $\text{B}_{\text{out}}$ is mixed on a 50-50 beamsplitter with a strong coherent state $\vert \delta_{\text{LO}} \rangle_{\text{C}}$  in mode C, which is called the local oscillator. The local oscillator may be taken from the same coherent radar source (CRS) used to perform the ranging in the monostatic Michelson configuration and it has a well define phase $\vartheta_\text{LO}$  so that $\big\vert \delta_\text{LO} \big\rangle_{\text{C}} = \big\vert \vert \delta_\text{LO} \vert \exp{(\mathrm{i} \vartheta_\text{LO})} \big\rangle_{\text{C}}$. It has an intensity $\bar{n}_\text{LO}=\big\vert \delta_\text{LO} \big\vert^2 \gg \bar{n}$, which is the origin of the amplification. The two outputs are then guided to the two detectors D and E where the intensity between them is differenced. That intensity difference, as a function of the unknown ranging phase $\varphi$, is our signal. There is a well-known but computationally intensive method for exploiting balanced homodyne detection to construct the Wigner function for an arbitrary quantum state, which is called quantum tomography \cite{Lvovsky2009}. However this process is slow and resource intensive, not something you want to do if you have a fast moving target, and so we take advantage of the fact that we know the output state is a coherent state and that we only need the Wigner function at one point, that is at the origin of phase space, and follow the simpler procedure outlined in Ref.~\cite{Plick2010}. Using the standard beamsplitter back-transformations, $\hat{e}\rightarrow1/\sqrt{2}(\hat{b}+\mathrm{i}\hat{c})$ and $\hat{d}\rightarrow 1/\sqrt{2}(\hat{c}+\mathrm{i}\hat{b})$ the intensity difference operator at the two detectors $\hat{\mathrm{M}}$ may be written as, $\hat{\mathrm{M}}=\hat{e}^\dagger\hat{e}-\hat{d}^\dagger\hat{d}=\mathrm{i}(\hat{b}^\dagger\hat{c}-\hat{c}^\dagger\hat{b})$. This allows us to compute the expectation of $\hat{\text{M}}$ in terms of the input states. From the states shown in Fig.~\ref{FIG04} we have
\begin{align}
Y_\text{B}(\vartheta_{\text{LO}},\varphi) 
\nonumber
& =~{_\text{B}}\big\langle \alpha_{\varphi} \big\vert ~{_\text{C}}\big\langle \delta_\text{LO} \big\vert \hat{\text{M}} \big\vert \delta_\text{LO} \big\rangle{_\text{C}} \big\vert \alpha_{\varphi} \big\rangle{_\text{B}} \\
\nonumber
& =~{_\text{B}}\big\langle \alpha_{\varphi} \big\vert ~{_\text{C}}\big\langle \delta_\text{LO} \big\vert \mathrm{i} 
\left(\hat{b}^\dagger\hat{c}-\hat{c}^\dagger\hat{b} \right) \big\vert \delta_\text{LO} \big\rangle{_\text{C}} \big\vert \alpha_{\varphi} \big\rangle{_\text{B}} \\
\nonumber
& =~\mathrm{i}\left( \langle \hat{b}^\dagger \rangle_\text{B} \langle \hat{c} \rangle_\text{C} - \langle \hat{c}^\dagger \rangle_\text{C} \langle \hat{b} \rangle_\text{B} \right)\\
\nonumber
& =~\mathrm{i}\left( \alpha_{\varphi}^* ~\vert \delta_\text{LO} \vert e^{\mathrm{i} \vartheta_\text{LO}} -\alpha_{\varphi} ~\vert \delta_\text{LO} \vert e^{-\mathrm{i} \vartheta_\text{LO}} \right) \\
\nonumber
& =~-2 \vert \alpha_{\varphi} \vert \vert \delta_\text{LO} \vert \sin(\vartheta_{\text{LO}}) \\
& =~-2 \sqrt{\bar{n}_\varphi} \sqrt{\bar{n}_\text{LO}} \sin(\vartheta_{\text{LO}}),
\label{EQN05}
\end{align}
where the fact $\alpha_{\varphi}^*=\alpha_{\varphi}$ is used and $\bar{n}_{\varphi}=\bar{n} \sin^2(\varphi/2)$ as before. A critical point to notice is that it is clear from Eq.~\eqref{EQN05} that a balanced quantum homodyne detector is an amplifier since, in general, $\bar{n}_\text{LO}\gg\bar{n}$. This amplification provides a critical advantage in that the amplified signal can be made to be well above the thermal electronic noise floor found even in the best radar detectors. Hence by using a quantum balanced homodyne approach the detectors need not be photon number resolving (difficult) but rather just highly efficient with a quantum efficiency approaching unity (easier) \cite{Leonhardt1997,Zmuidzinas2004}. In addition, the intensity differencing in balance homodyne detection removes all technical noise and classical noise of the local oscillator. Finally, a well-known result in coherent LIDAR, the signal-to-noise of the output is limited by the shot-noise of the local oscillator that scales like $\sqrt{\bar{n}_\text{LO}}$ \cite{Belmonte2010}. 

On the other hand, from Eq.~\eqref{EQN05} we have
\begin{align}
\vert \alpha_{\varphi} \vert = -\frac{Y_\text{B}(\vartheta_{\text{LO}},\varphi)}{2 \sqrt{\bar{n}_\text{LO}} \sin(\vartheta_{\text{LO}})}.
\label{EQN06}
\end{align}
This can be substituted into the general expressions Eqs.~\eqref{EQN01} and \eqref{EQN02} and the parity signal from mode $\text{B}_{\text{out}}$ becomes
\begin{align}
\langle \hat{\mathrm{\Pi}} \rangle = \frac{\pi}{2}W(0,0)=\exp{(-2\vert \alpha_\varphi \vert^2)}.
\label{EQN07}
\end{align} 
Setting the phase of the local oscillator to $\vartheta_{\text{LO}}=\pi/2$  we have that our reconstructed parity signal is, 
\begin{align}
S_{\text{B}}(\varphi)=~{_\text{B}}\big\langle \alpha_{\varphi}\big\vert \hat{\mathrm{\Pi}} \big\vert \alpha_{\varphi} \big\rangle{_\text{B}}
=\exp \left(-\frac{Y_{\text{B}}^{2}(\pi/2,\varphi)}{2\bar{n}_\text{LO}}\right),
\label{EQN08}
\end{align}
where we emphasize that $Y_{\text{B}}^{2}(\pi/2,\varphi)$ \textit{is the measured (normalized) intensity difference between detectors D and E}. By combining Eq.~\eqref{EQN05} and ~\eqref{EQN08} we recover the parity result in Eq.~\eqref{EQN03}. However by going the route of balanced homodyne detection we have gained control over signal to noise and have done away for the need for photon-number resolving radar detectors. As long as the detectors have a high quantum efficiency, the balanced quantum homodyne scheme is capable of detecting even a return signal with $\bar{n}_{\varphi}=1$, that is where on average only one radar photon is present \cite{Leonhardt1997}. 

Since every photon is precious we should not ignore the upper exit port $\text{A}_{\text{out}}$, in Fig.~\ref{FIG02}, where if we are working near the ‘sweet spot’ of $\varphi \cong 0$  is where most of the signal photons will emerge. Since only phase differences and not absolute phases have meaning, whenever we talk about measuring $\varphi$ we really mean that we are measuring the phase difference between the phase in the target $\varphi_{\text{T}}$ arm and that in the reference arm $\varphi_{\text{R}}$ of the interferometer. For simplicity we have set the phase of the reference arm to zero in which case $\varphi=\varphi_{\text{T}}-0$ is indeed that phase difference. In actual operation one would put a tunable phase shifter in the reference arm, with phase difference $\varphi=\varphi_{\text{T}}-\varphi_{\text{R}}$ being the signal. In this way, typically in a feed-back loop, as we gather information about $\varphi_{\text{T}}$ in the data we can tune the interferometer in real time to always maintain the ‘sweet spot’ condition $\varphi\cong0$. This tuning also gives us information about the absolute phase difference. The problem is that tuning the signal in the lower output port $\text{B}_{\text{out}}$ to the sweet spot moves it to a phase point where the signal in the upper port $\text{A}_{\text{out}}$ is not super resolved. This can be fixed by deploying a slightly different homodyne technique at the upper port. From the tuning in the reference arm and the signal in the lower port we will have enough information about the unknown range phase $\varphi$ to do the following. In the homodyne measurement at $\text{A}_{\text{out}}$ we take the phase of the local oscillator to be $\vartheta_{\text{LO}}\cong\varphi/2$, in which case Eq.~\eqref{EQN05} becomes, $Y_{\text{A}}(\varphi/2,\varphi)=\sqrt{\bar{n}_\text{LO}}\sqrt{\bar{n}}\sin(\varphi)$, from which we can extract, 
\begin{align}
S_{\text{A}}(\varphi)
\nonumber
& =\exp \left(-\frac{Y_{\text{A}}^{2}(\varphi/2,\varphi)}{2\bar{n}_\text{LO}} \right) \\
\nonumber
& =\exp \left(-\bar{n}\sin^2(\varphi)/2 \right) \\
& \cong \exp(-\bar{n}\varphi^2/2),
\label{EQN09}
\end{align}
where the last term is taken near $\varphi\cong 0$ and so the range resolution is the same and that of Eq.~\eqref{EQN04} again. The outputs signals of the two ports are then simply averaged to give the best estimate of the range phase. 

\section{SUPER-RESOLVED ANGLE DETERMINATION}
\begin{figure}
	\centering
		 \includegraphics[width=0.5\textwidth]{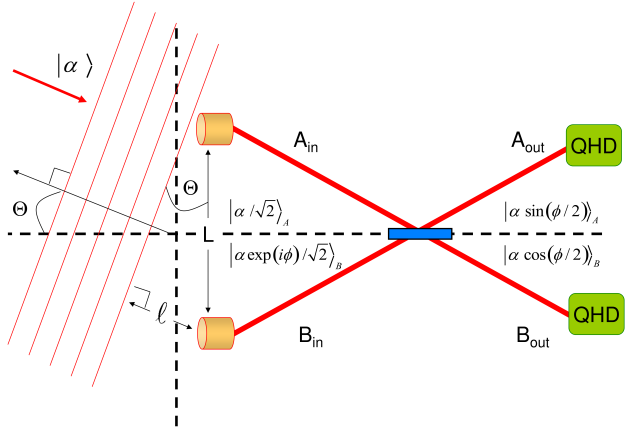}
	\caption{Here we indicate how to use quantum homodyne detection to extract the super-resolved altitudinal angle $\varTheta$. Two resonant radar cavity detectors are placed a distance $L$ apart and connected via a 50-50 beam splitter (balanced mixer) as shown. The incoming coherent signal $\vert \alpha \rangle$ arrives as a plane wave and hence the state at the lower cavity experiences a phase shift $\phi = k \ell$ with respect to that at the upper cavity due to the path difference $\ell = L\sin(\varTheta)$. (Here the wave number is $k = 2 \pi /\lambda$.) As before the two signals mix at the beam splitter and quantum homodyne detection (QHD) is performed at each output providing a super-resolved measurement of phase with resolution $\delta\phi=1/\sqrt{\bar{n}}$. Since $L$ is fixed and known this yields a super-resolved measurement of the altitudinal angle with a resolution approximately equal to $\delta \varTheta_{\text{Q}} =  \lambda / 2 \pi (L \sqrt{\bar{n}})$ for a target close to the horizon. This is a factor of $2 \pi \sqrt{\bar{n}}$ smaller than the classical diffraction limit of $\delta \varTheta_{\text{C}} = \lambda / L$.}
	\label{FIG05}
\end{figure}

In the preceding section we described how to obtain super-resolved ranging information using monostatic Michelson interferometer combined with quantum homodyne detection. Critical for complete target location is super-resolved angle information as well. Particularly this is difficult to obtain at radar wavelengths since, as discussed in the introduction, the return signal arrives as a plane wave. We discuss here how to get such a signal for the altitudinal angle $\varTheta$ and azimuthal angle $\varPhi$. The insight is to realize that a MZI can be mapped onto a two-slit diffraction configuration and thus the desired unknown angle can be mapped into an unknown phase and then we use the same homodyne technique to measure that phase and hence extract the angle.  Consider in Fig.~\ref{FIG05} the return signal arriving at the detector as a plane wave with a Poynting vector at an angle $\varTheta$ with respect to the horizon. We may treat the signal again as a coherent state $\vert \alpha \rangle$ where any phase accumulated on the journey to the target and back is suppressed for this current discussion. Two resonant receiver cavities are placed a distance $L$ apart as shown and connected by wave-guides to a balanced mixer (beamsplitter) and then we perform quantum homodyne at each output as before. The coherent state will be split over the two receivers but the lower state acquires a relative phase shift $\phi=k\ell$ with respect to that at the upper cavity due to the path difference $\ell = L \sin(\varTheta)$. (Here the wavenumber is $k = 2 \pi /\lambda$ .) From this point on the measurement of that phase shift is carried out precisely as before, as in Fig.~\ref{FIG04}, by performing quantum homodyne measurement at the two outputs. (The output states are reversed here since the phase accumulated is in the lower rather than in the upper arm.) The resolution of the phase is again $\delta\phi=1/\sqrt{\bar{n}}$ for $\phi \cong 0$ corresponding to $\varTheta \cong 0$ that is a target close to the horizon. 

Since $L$ is fixed and known this yields a super-resolved measurement of the altitudinal angle with a resolution approximately equal to,
\begin{align}
\delta \varTheta_{\text{Q}} \cong \frac{\lambda}{L}\frac{1}{2\pi \sqrt{\bar{n}}},
\label{EQN10}
\end{align}
close to the horizon. This is a factor of $2 \pi \sqrt{\bar{n}}$ smaller than the classical diffraction limit of $\delta \varTheta_{\text{C}}=\lambda/L$. For $\bar{n}=100$  average return photons the quantum result is 60 times smaller than the classical diffraction limit. Rotating the detector system about its bilateral axis allows us to extract the azimuthal phase $\varPhi$ with the same resolution. These two angular measurements, when combined with the range measurement, completely characterize the instantaneous location of the target. Temporal differentiation of the signal then is performed to acquire the velocity and acceleration. 

\section{SUMMARY AND CONCLUSIONS}
In this paper we have presented an outline to implement a super-resolving radar ranging and angular determination system that utilizes a quantum parity detection scheme implemented with a homodyne detection process. We have argued that a similar quantum on-off detection scheme, also implemented with homodyne detection, gives identical super-resolving, sub-Rayleigh-diffraction-limited performance \cite{Distante2013}. In quantum metrology the discussions of super-sensitivity (signal-to-noise) below the shot-noise limit is often discussed in the same context of super-resolution (sub-Rayleigh) below the diffraction limit \cite{Dowling2008}. The reason that these two properties are closely related is that the slopes of the detected interferogram in Fig.~\ref{FIG03} determines directly the resolution, while the same slope appears in the formula that is used to estimate phase sensitivity.

\textit{In the absence of loss}, it is well known that the protocol for producing maximal phase sensitivity (sub-shot-noise at the Heisenberg limit) and simultaneously super-resolution (sub-Rayleigh features $N$ times smaller than the wavelength) is to transmit an entangled state of the electromagnetic field, called a N00N state, in an interferometric set up as shown in Fig.~\ref{FIG01}(a) \cite{Durkin2007}. However, as discussed in Section I, several groups have shown that, in the presence of high loss (due to absorption, scattering, and diffraction), the optimal strategy is to transmit coherent states to the target, in which case the sensitivity can be at best only at shot-noise in the return signal. Those previous works left open the question, is super-resolution possible at the shot-noise limit? Our answer to this question is yes, and two proof-of-principle optical experiments back up our assertion \cite{Distante2013, Eisenberg2012}. 

What we have shown in this present work is that, setting super-sensitivity aside as a goal, that it is possible with coherent state sources and a homodyne implementation of a quantum parity detection scheme to beat the Rayleigh resolution by an arbitrary amount \cite{Gao2010}. Distante \emph{et al.} showed that on-off detection implemented with homodyne shows identical results \cite{Distante2013}. Even better, this super-resolution is obtained while maintaining sensitivity at shot-noise, which is probably the best sensitivity attainable with coherent states, and which utilizes all the returned photons. We reviewed in Section II that super-resolved ranging is attainable, in Section III how to extract this with homodyne, and then we showed in Section IV that altitudinal and azimuthal angular determination is also possible using a modified homodyne scheme. Given that ordinary radar systems already transmit coherent states of the electromagnetic field, and also given that homodyne detection is a standard radar detection technique (balanced mixer radiometry) our entire proposed quantum radar system can be implemented mostly with commercial-off-the-shelf components with minimal redesign of existing radar systems. 

In the context of quantum imaging and metrology the question arises as to what is the relevant figure of merit when designing a practical system, resolution or sensitivity? These two properties of the system are closely related in that the slope of a super-resolving interference fringe in part determines the sensitivity of the device \cite{Dowling2008}. For example in coherent quantum lithography, resolution is the only relevant figure of merit, and discussions of sensitivity never occur. In lithography the Rayleigh diffraction limit rules \cite{Boto2000, Kok2001}. The goal in lithography is to place features as close together as possible, closer than the wavelength of light, and sensitivity is never discussed. However, in contrast, in the Laser Interferometer Gravitational Wave Observatory (LIGO), they can do many orders of magnitude better than the Rayleigh diffraction limit in distance measurements, and that community never discusses resolution and only concerns themselves with sensitivity, which currently is at shot-noise for a large range of frequencies. 

The LIGO interferometers have a circulating laser power of 100 kW and they can measure relative arm displacements on the order of an attometer while deploying laser radiation of a wavelength on the order of a micron. That is they are doing 12 orders of magnitude better than the Rayleigh diffraction limit. A radar system designer might ask how that is possible. The answer has three parts. Firstly, in LIGO the gravity wave causes the arms of the interferometer to change very slowly. Most of the time the gravity wave is not present and so they can lock the interferometer. Secondly they have huge numbers of photons to work with, about $\bar{n}\cong 10^{20}$  per second in the interferometer. Thirdly they have the luxury to integrate their data over periods of hours, days, and even weeks. In this manner they beat down the noise until the minimal detectible arm displacement is given by the quantum shot-noise expression $\Delta x = \lambda/(2 \pi \sqrt{\bar{n}})$ , which gives the attometer precision. 

Why is a radar ranging system not like LIGO? The answer is that, for many applications, the radar operators do not have the luxury to integrate their data over long periods of time. Recall that target speeds of Mach 30 are not uncommon for some applications. Hence the interference fringes are moving very rapidly. There is no hope to lock down the interferometer and integrate at these speeds. In addition data integration times are measured in seconds, not hours or days. And finally the number of photons arriving at the detector is very small so there is little data to integrate. The strategy then for coherent lidar and radar is to track the narrowest feature you can find in the interferogram and follow this to establish the range and angular position parameter $R$, $\varTheta$ and $\varPhi$, and then temporally differentiate these in real time to extract vector velocity and acceleration. It is for these reasons that radar designers often only worry about diffraction and seldom about sensitivity. It is in such a scenario that we propose the use of our super-resolving quantum radar scheme.
 
In this paper as well as in related works \cite{Gao2010, Distante2013, Eisenberg2012}, we are using the term `super resolution' in a slightly different fashion than in previous works. In previous work on quantum lithography, super resolution was used to mean many fringes per unit wavelength \cite{Boto2000, Kok2001}. This usage explicitly implied narrower fringes in that more fringes per wavelength necessarily implies narrower ones. In this current paper we restrict the usage to only the narrowing of the fringes. It is clear from Fig.~\ref{FIG03} that the spacing between the fringes is still at the classical wavelength, giving our interferogram the look of a typical Fabry-P$\acute{\text{e}}$rot output with a free-spectral range of one wavelength. For coherent interferometric lithography, the increased number of fringes per wavelength is critical to the capability of writing more features $N$-times closer than is possible classically. However the increased fringe number per wavelength is not so critical in radar ranging. For radar what is important is that, once one has locked on to a particular fringe, one can tell if the fringe has moved and if so by how much. That sets the resolution, particularly on a rapidly changing range $R$. The one-dimensional Rayleigh criterion then holds --- one can tell if the fringe moves by one full width at half maximum. From Fig.~\ref{FIG03} we see that this distance is classically $\delta x_{\text{C}} = \lambda$ but that in our quantum scheme proposed here it is $\delta x_{q} = \lambda/(2 \pi \sqrt{\bar{n}})$, which is approximately 60 times narrower than the classical result for a return power of photons. It is interesting to note that, since our scheme is also shot-noise limited, that the minimum sensitivity has the same scaling, namely, $\Delta x_{q} = \lambda/(2 \pi \sqrt{\bar{n}})$. However, as we have argued above, in a situation where the target moves quickly, there is little time to integrate data, and the return number of photons is small, that the resolution and not sensitivity is the relevant metric of system performance. One of the most important points to notice for our proposed scheme is that we have mapped this $1/\sqrt{\bar{n}}$ scaling out of the sensitivity (where it is useless for our application) and into the resolution (where it is critical). 

In principle, for complete ranging, we need to have not only narrow fringes but in addition we need to know which fringe we are on. Without the latter information we can specify $R$ only modulo a wavelength. This ‘which-fringe’ information can be obtained by using a standard technique in radar ranging: We simply apply a narrow temporal chirp in the outgoing radar beam and time its round trip to the target and back. That then gives us an absolute distance measurement to supply the needed information as to what fringe we are locked on and hence completely determines $R$. (In practical radar systems one must compensate for changes in the atmospheric index of refraction as a function of distance, altitude, weather, time of day, etc. Such models of the index are well developed and may be deployed here in our scheme with little or no change.)  

Finally, even though we have only considered the use of the ‘quantum’ radar system when the return power is very small, our proposal is likely to be useful even at high return powers --- the quadrature noise measured in homodyne detection is independent of the excitation of the coherent state. Moreover, in that high-power regime the entire quantum radar system may described within the context of classical radar theory. In the high-return-power regime a better term for our system may be a `quantum-inspired' radar. 

This term quantum inspired has an interesting recent history where researchers mimicked truly quantum electromagnetic field performance improvements by using a non-intuitive classical electromagnetic formalism inspired from the quantum setup. A good example of a quantum inspired classical technology has recently been demonstrated in the group of Kevin Resch in the form of dispersion cancellation in optical coherence tomography --- a type of interferometric microscopy. In 1992 James Franson, as well as the group of Raymond Chiao, pointed out that it was possible to cancel second-order dispersion using non-classical light from parametric down conversion, a source of entangled photons \cite{Franson1992, Steinberg1992}. Abouraddy \emph{et al.} applied this technique to improve the resolution in quantum coherence tomography microscopy where dispersion blurs the images \cite{Abouraddy2002, Booth2011}. Resch and collaborators were able to show that they could classically mimic the quantum dispersion cancellation effect by engineering classical chirped and anti-chirped light pulses whose spectral characteristics closely matched those of the quantum light source \cite{Kaltenbaek2008}. In a similar fashion, Jeffery Shapiro and collaborators were also able to design a ‘quantum inspired’ dispersion canceling optical coherence tomography system utilizing nonlinear (but classical) optics in the form of optical phase conjugation \cite{LeGouet2010}. In both cases the classical systems did not exist until they were constructed in analogy to the quantum systems. In this sense our quantum radar scheme proposed may well turn out to be a non-intuitive classical but quantum-inspired system in the high-return-power regime. 

In summary, we have presented a quantum radar system with super-resolving ranging and angular determination that is much below the classical Rayleigh diffraction limit. The system will be particularly useful for the radar tracking of far distant and fast moving objects in which little radar power returns to the detector. The system can be implemented using mostly off-the-shelf technologies with only minor modifications to current radar systems.

\section*{ACKNOWLEDGEMENTS}
K. J. and J. P. D. would like to acknowledge support from the National Science Foundation and the Air Force Office of Scientific Research. This work is also supported by the Intelligence Advanced Research Projects Activity (IARPA) via Department of Interior National Business Center contract number D12PC00527. The U.S. Government is authorized to reproduce and distribute reprints for Governmental purposes notwithstanding any copyright annotation thereon. Disclaimer: The views and conclusions contained herein are those of the authors and should not be interpreted as necessarily representing the official policies or endorsements, either expressed or implied, of IARPA, DoI/NBC, or the U.S. Government.


%

\end{document}